\documentclass[12pt]{article}

\usepackage[T1]{fontenc}

\usepackage{graphicx, amsmath, amsthm, amssymb,xcolor,tikz, import}
\usepackage{authblk}
\usepackage[hidelinks]{hyperref}
\usepackage{caption}
\theoremstyle{definition}

\newtheorem{prop}{Proposition}
\newtheorem{lemma}{Lemma}

\usepackage[margin=1in]{geometry}
\usepackage{xcolor}

\usepackage{setspace}
\newcommand{\nc}{\newcommand}

\numberwithin{equation}{section}
\allowdisplaybreaks

\nc{\renc}{\renewcommand}
\nc{\ssec}{\subsection}
\nc{\sssec}{\subsubsection}
\nc{\on}{\operatorname}

\nc{\ips}{{\iota_P^{(S)}}}
\nc{\ipms}{{\iota_{P^-}^{(S)}}}
\nc{\sfpps}{{\sfp_P^{(S)}}}
\nc{\sfppms}{{\sfp_{P^-}^{(S)}}}

\nc\ol{\overline}
\nc\wt{\widetilde}
\nc\tboxtimes{\wt{\boxtimes}}
\nc\tstar{\wt{\star}}
\nc{\alp}{\alpha}

\nc{\ZZ}{{\mathbb Z}}
\nc{\NN}{{\mathbb N}}
\nc{\BF}{{\mathbb F}}
\nc{\OO}{{\mathbb O}}
\renc{\SS}{{\mathbb S}}
\nc{\DD}{{\mathbb D}}
\nc{\GG}{{\mathbb G}}

\nc{\Fq}{{\mathbb F}_q}
\nc{\Fqb}{\ol{\mathbb F}_q}
\nc{\Ql}{{\mathbb Q}_\ell}
\nc{\Qlb}{{\ol{\mathbb Q}_\ell}}
\nc{\id}{\text{id}}
\nc\X{\mathcal X}

\nc{\red}{\on{red}}
\nc{\Ho}{\on{Ho}}
\nc{\Hom}{\on{Hom}}
\nc{\coef}{\on{coef}}
\nc{\Lie}{\on{Lie}}
\nc{\Diff}{\on{Diff}}
\nc{\Loc}{\on{Loc}}
\nc{\Pic}{\on{Pic}}
\nc{\Bun}{\on{Bun}}
\nc{\IC}{\on{IC}}
\nc{\Aut}{\on{Aut}}
\nc{\rk}{\on{rk}}
\nc{\Sh}{\on{Sh}}
\nc{\Perv}{\on{Perv}}
\nc{\pos}{{\on{pos}}}
\nc{\Conv}{\on{Conv}}
\nc{\Sph}{\on{Sph}}
\nc{\Sym}{\on{Sym}}
\nc{\BunBb}{\overline{\Bun}_B}
\nc{\BunNb}{\overline{\Bun}_N}
\nc{\BunTb}{\overline{\Bun}_T}
\nc{\BunBbm}{\overline{\Bun}_{B^-}}
\nc{\BunBbel}{\overline{\Bun}_{B,el}}
\nc{\BunBbmel}{\overline{\Bun}_{B^-,el}}
\nc{\Buno}{\overset{o}{\Bun}}
\nc{\BunPb}{{\overline{\Bun}_P}}
\nc{\BunBM}{\Bun_{B(M)}}
\nc{\BunBMb}{\overline{\Bun}_{B(M)}}
\nc{\BunPbw}{{\widetilde{\Bun}_P}}
\nc{\BunBP}{\widetilde{\Bun}_{B,P}}
\nc{\GUb}{\overline{G/U}}
\nc{\GUPb}{\overline{G/U(P)}}

\nc{\Hhom}{\underline{\on{Hom}}}
\nc\syminfty{\on{Sym}^{\infty}}
\nc\lal{\ol{\lambda}}
\nc\xl{\ol{x}}
\nc\thl{\ol{\theta}}
\nc\nul{\ol{\nu}}
\nc\mul{\ol{\mu}}
\nc\Sum{$\Sigma$}
\nc{\oX}{\overset{\circ}{X}{}}
\nc{\wtoX}{\wt{\overset{\circ}{X}}{}}
\nc{\hl}{\overset{\leftarrow}h{}}
\nc{\hr}{\overset{\rightarrow}h{}}
\nc{\M}{{\mathcal M}}
\nc{\N}{{\mathbb N}}
\nc{\F}{{\mathcal F}}
\nc{\D}{{\mathcal D}}
\nc{\Q}{{\mathcal Q}}
\nc{\Y}{{\mathcal Y}}
\nc{\G}{{\mathcal G}}
\nc{\E}{{\mathcal E}}
\nc{\CalC}{{\mathcal C}}
\nc\Dh{\widehat{\D}}

\nc{\C}{{\mathcal C}}
\nc{\K}{{\mathcal K}}
\renewcommand{\H}{{\mathcal H}}

\nc{\T}{{\mathcal T}}
\nc{\V}{{\mathcal V}}
\renc{\P}{{\mathcal P}}
\nc{\A}{{\mathcal A}}
\nc{\B}{{\mathcal B}}
\nc{\U}{{\mathcal U}}

\nc{\Gr}{{\on{Gr}}}

\nc{\frn}{{\check{\mathfrak u}(P)}}

\nc{\fC}{\mathfrak C}
\nc{\p}{\mathfrak p}
\nc{\q}{\mathfrak q}
\nc\f{{\mathfrak f}}

\nc{\qo}{{\mathfrak q}}
\nc{\po}{{\mathfrak p}}
\nc{\s}{{\mathfrak s}}
\nc\w{\text{w}}

\nc\Spec{\on{Spec}}
\nc\Proj{\on{Proj}}
\nc\Mod{\on{Mod}}
\nc{\tw}{\widetilde{\mathfrak t}}
\nc{\pw}{\widetilde{\mathfrak p}}
\nc{\qw}{\widetilde{\mathfrak q}}
\nc{\jw}{\widetilde j}

\nc{\grb}{\overline{\Gr}}
\nc{\I}{\mathcal I}

\nc{\lambdach}{{\check\lambda}}
\nc{\Lambdach}{{\check\Lambda}{}}
\nc{\much}{{\check\mu}}
\nc{\omegach}{{\check\omega}}
\nc{\nuch}{{\check\nu}}
\nc{\etach}{{\check\eta}}
\nc{\alphach}{{\check\alpha}}
\nc{\oblvtach}{{\check\oblvta}}
\nc{\rhoch}{{\check\rho}}
\nc{\ch}{{\check h}}

\nc{\Hb}{\overline{\H}}

\nc{\BA}{{\mathbb{A}}}
\nc{\BC}{{\mathbb{C}}}
\nc{\BG}{{\mathbb{G}}}
\nc{\BM}{{\mathbb{M}}}
\nc{\BO}{{\mathbb{O}}}
\nc{\BD}{{\mathbb{D}}}
\nc{\BN}{{\mathbb{N}}}
\nc{\BP}{{\mathbb{P}}}
\nc{\BQ}{{\mathbb{Q}}}
\nc{\BR}{{\mathbb{R}}}
\nc{\BZ}{{\mathbb{Z}}}
\nc{\BS}{{\mathbb{S}}}
\nc{\Deep}{{\bf{deep}}}
\nc{\deep}{deep}

\renewcommand{\BP}{{\mathbb{P}}}
\nc{\BE}{{\mathbb{E}}}

\nc{\CA}{{\mathcal{A}}}
\nc{\CB}{{\mathcal{B}}}

\nc{\CE}{{\mathcal{E}}}
\nc{\CF}{{\mathcal{F}}}
\nc{\CH}{{\mathcal{H}}}

\nc{\CL}{{\mathcal{L}}}
\nc{\CC}{{\mathcal{C}}}
\nc{\CG}{{\mathcal{G}}}
\nc{\CalD}{{\mathcal{D}}}
\nc{\cD}{{\mathcal{D}}}
\nc{\CM}{{\mathcal{M}}}
\nc{\CN}{{\mathcal{N}}}
\nc{\CK}{{\mathcal{K}}}
\nc{\CO}{{\mathcal{O}}}
\nc{\CP}{{\mathcal{P}}}
\nc{\CQ}{{\mathcal{Q}}}
\nc{\CR}{{\mathcal{R}}}
\nc{\CS}{{\mathcal{S}}}
\nc{\CT}{{\mathcal{T}}}
\nc{\CU}{{\mathcal{U}}}
\nc{\CV}{{\mathcal{V}}}
\nc{\CW}{{\mathcal{W}}}
\nc{\CX}{{\mathcal{X}}}
\nc{\CY}{{\mathcal{Y}}}
\nc{\CZ}{{\mathcal{Z}}}
\nc{\CI}{{\mathcal{I}}}

\nc{\csM}{{\check{\mathcal A}}{}}
\nc{\oM}{{\overset{\circ}{\mathcal M}}{}}
\nc{\obM}{{\overset{\circ}{\mathbf M}}{}}
\nc{\oCA}{{\overset{\circ}{\mathcal A}}{}}
\nc{\obA}{{\overset{\circ}{\mathbf A}}{}}
\nc{\ooM}{{\overset{\circ}{M}}{}}
\nc{\osM}{{\overset{\circ}{\mathsf M}}{}}
\nc{\vM}{{\overset{\bullet}{\mathcal M}}{}}
\nc{\nM}{{\underset{\bullet}{\mathcal M}}{}}
\nc{\oD}{{\overset{\circ}{\mathcal D}}{}}
\nc{\obD}{{\overset{\circ}{\mathbf D}}{}}
\nc{\oA}{{\overset{\circ}{\mathbb A}}{}}
\nc{\op}{{\overset{\bullet}{\mathbf p}}{}}
\nc{\cp}{{\overset{\circ}{\mathbf p}}{}}
\nc{\oU}{{\overset{\bullet}{\mathcal U}}{}}
\nc{\oZ}{{\overset{\circ}{\mathcal Z}}{}}
\nc{\ofZ}{{\overset{\circ}{\mathfrak Z}}{}}
\nc{\oF}{{\overset{\circ}{\fF}}}
\nc{\oC}{{\overset{\circ}{\on{C}}}}

\nc{\fa}{{\mathfrak{a}}}
\nc{\fb}{{\mathfrak{b}}}
\nc{\fd}{{\mathfrak{d}}}
\nc{\ff}{{\mathfrak{f}}}
\nc{\fg}{{\mathfrak{g}}}
\nc{\fgl}{{\mathfrak{gl}}}
\nc{\fh}{{\mathfrak{h}}}
\nc{\fj}{{\mathfrak{j}}}
\nc{\fk}{{\mathfrak{k}}}
\nc{\fl}{{\mathfrak{l}}}
\nc{\fm}{{\mathfrak{m}}}
\nc{\fn}{{\mathfrak{n}}}
\nc{\fu}{{\mathfrak{u}}}
\nc{\fo}{{\mathfrak{o}}}
\nc{\fp}{{\mathfrak{p}}}
\nc{\fr}{{\mathfrak{r}}}
\nc{\fs}{{\mathfrak{s}}}
\nc{\fso}{{\mathfrak{so}}}
\nc{\fsp}{{\mathfrak{sp}}}
\nc{\ft}{{\mathfrak{t}}}
\nc{\fsu}{{\mathfrak{su}}}
\nc{\fv}{{\mathfrak{v}}}
\nc{\fz}{{\mathfrak{z}}}
\nc{\fsl}{{\mathfrak{sl}}}
\nc{\hsl}{{\widehat{\mathfrak{sl}}}}
\nc{\hgl}{{\widehat{\mathfrak{gl}}}}
\nc{\hg}{{\widehat{\mathfrak{g}}}}
\nc{\chg}{{\widehat{\mathfrak{g}}}{}^\vee}
\nc{\hn}{{\widehat{\mathfrak{n}}}}
\nc{\chn}{{\widehat{\mathfrak{n}}}{}^\vee}

\nc{\fA}{{\mathfrak{A}}}
\nc{\fB}{{\mathfrak{B}}}
\nc{\fD}{{\mathfrak{D}}}
\nc{\fE}{{\mathfrak{E}}}
\nc{\fF}{{\mathfrak{F}}}
\nc{\fG}{{\mathfrak{G}}}
\nc{\fK}{{\mathfrak{K}}}
\nc{\fL}{{\mathfrak{L}}}
\nc{\fM}{{\mathfrak{M}}}
\nc{\fN}{{\mathfrak{N}}}
\nc{\fP}{{\mathfrak{P}}}
\nc{\fU}{{\mathfrak{U}}}
\nc{\fV}{{\mathfrak{V}}}
\nc{\fZ}{{\mathfrak{Z}}}

\nc{\ba}{{\mathbf{a}}}
\nc{\bb}{{\mathbf{b}}}
\nc{\bc}{{\mathbf{c}}}
\nc{\bd}{{\mathbf{d}}}
\nc{\bbf}{{\mathbf{f}}}
\nc{\be}{{\mathbf{e}}}
\nc{\bi}{{\mathbf{i}}}
\nc{\bj}{{\mathbf{j}}}
\nc{\bn}{{\mathbf{n}}}
\nc{\bo}{{\mathbf{o}}}
\nc{\bp}{{\mathbf{p}}}
\nc{\bq}{{\mathbf{q}}}
\nc{\br}{{\mathbf{r}}}
\nc{\bu}{{\mathbf{u}}}
\nc{\bv}{{\mathbf{v}}}
\nc{\bx}{{\mathbf{x}}}
\nc{\bs}{{\mathbf{s}}}
\nc{\bt}{{\mathbf{t}}}
\nc{\by}{{\mathbf{y}}}
\nc{\bw}{{\mathbf{w}}}
\nc{\bA}{{\mathbf{A}}}
\nc{\bK}{{\mathbf{K}}}
\nc{\bB}{{\mathbf{B}}}
\nc{\bC}{{\mathbf{C}}}
\nc{\bG}{{\mathbf{G}}}
\nc{\bD}{{\mathbf{D}}}
\nc{\bH}{{\mathbf{H}}}
\nc{\bM}{{\mathbf{M}}}
\nc{\bN}{{\mathbf{N}}}
\nc{\bO}{{\mathbf{O}}}
\nc{\bV}{{\mathbf{V}}}
\nc{\bW}{{\mathbf{W}}}
\nc{\bX}{{\mathbf{X}}}
\nc{\bZ}{{\mathbf{Z}}}
\nc{\bS}{{\mathbf{S}}}

\nc{\sA}{{\mathsf{A}}}
\nc{\sB}{{\mathsf{B}}}
\nc{\sC}{{\mathsf{C}}}
\nc{\sD}{{\mathsf{D}}}
\nc{\sF}{{\mathsf{F}}}
\nc{\sG}{{\mathsf{G}}}
\nc{\sK}{{\mathsf{K}}}
\nc{\sk}{{\mathsf{k}}}
\nc{\sM}{{\mathsf{M}}}
\nc{\sO}{{\mathsf{O}}}
\nc{\sW}{{\mathsf{W}}}
\nc{\sQ}{{\mathsf{Q}}}
\nc{\sP}{{\mathsf{P}}}
\nc{\sR}{{\mathsf{R}}}
\nc{\sZ}{{\mathsf{Z}}}
\nc{\sfp}{{\mathsf{p}}}
\nc{\sfq}{{\mathsf{q}}}
\nc{\sr}{{\mathsf{r}}}
\nc{\bk}{{\mathsf{k}}}
\nc{\sg}{{\mathsf{g}}}
\nc{\sff}{{\mathsf{f}}}
\nc{\sfb}{{\mathsf{b}}}
\nc{\sfc}{{\mathsf{c}}}
\nc{\sd}{{\mathsf{d}}}

\nc{\var}{\text{Var}}

\newtheorem{corollary}{Corollary}
\newtheorem{remark}{Remark}

\usepackage{caption}
\usepackage{hyperref}
\usepackage{algorithm}
\usepackage{algpseudocode}
\usepackage{amsmath}

\begin{document}

\title{Analyzing Generalized Pólya Urn Models using Martingales, with an Application to Viral Evolution}
\author[1]{Ivan Specht*}
\author[2]{Michael Mitzenmacher}
\affil[1]{Harvard College, Faculty of Arts and Sciences, Harvard University, Cambridge, MA 02138. ORCiD: 0000-0003-2834-8191.}
\affil[2]{Department of Computer Science, School of Engineering and Applied Sciences, Harvard University, Cambridge, MA 02138. ORCiD: 0000-0001-5430-5457.}
\affil[*]{For correspondence: \texttt{ispecht@college.harvard.edu}.}
\maketitle

\newpage

\begin{abstract}
    \noindent The randomized play-the-winner (RPW) model is a generalized Pólya Urn process with broad applications ranging from clinical trials to molecular evolution. We derive an exact expression for the variance of the RPW model by transforming the Pólya Urn process into a martingale, correcting an earlier result of Matthews and Rosenberger (1997). We then use this result to approximate the full probability mass function of the RPW model for certain parameter values relevant to genetic applications. Finally, we fit our model to genomic sequencing data of SARS-CoV-2, demonstrating a novel method of estimating the viral mutation rate that delivers comparable results to existing scientific literature.

    \smallskip

\end{abstract}

\section{Introduction}

Consider the following generalized Pólya Urn model: An urn starts out with $u$ white balls and $v$ black balls, with $u + v > 0$. At each step $i=1, 2, 3, \dots$, a ball in the urn is chosen uniformly at random. If the chosen ball is white, a black ball is added to the urn with probability $p_W$ and a white ball is added with probability $1-p_W$, where $0 < p_W < 1$. If the chosen ball is black, a white ball is added to the urn with probability $p_B$ and a white ball is added with probability $1-p_B$, where $0 < p_B < 1$. The originally chosen ball is then returned to the urn, so that at step $i$, the total number of balls equals $u + v + i$. This model, known as the randomized play-the-winner (RPW) model, has been widely studied in  theoretical and applied contexts, with applications ranging from clinical trials to genetic mutations. Several papers, including Wei and Durham (1978), Smythe and Rosenberger (1995), Smythe (1996), and Rosenberger and Sriram (1997) have considered the RPW model's asymptotic properties, deriving limit theorems related to the asymptotic fraction of white (or black) balls in the urn.


The distribution of this fraction after finitely many steps, however, is less well understood. Rosenberger and Sriram (1997) prove an expression for its expectation, which we re-derive using an alternative approach. Matthews and Rosenberger (1997) propose an expression for its variance; however, we claim that this expression is erroneous, as it returns the incorrect variance for the special case $p_W = p_B = \frac{1}{2}$, which reduces the RPW model to the Binomial$(\frac{1}{2})$ model (see Appendix). In this paper, we introduce a new approach to compute the variance of the number of white balls in the RPW model after finitely many steps. Our method involves rewriting the RPW model as a martingale process---a transformation that may be applied not only to the RPW model, but to any process represented by a sequence of random variables $M_0, M_1, M_2, \dots$ with finite first and second moments such that $\BE[M_{i+1} | M_i]$ is a non-constant linear function of $M_i$ and $\var[M_{i+1} | M_i]$ is a quadratic function of $M_i$, for all $i$. We then use our variance formula to approximate the full probability mass function (PMF) of $M_i$ when $p_W$ and $p_B$ are small. For such $p_W$, $p_B$, the RPW model aptly characterizes branching processes that arise in viral genetics, with the balls representing viral particles and colors representing variants. We conclude with a novel method of estimating mutation rates of pathogens by fitting our approximate PMF to genome sequencing data of viruses.

\section{Constructing the Martingale}

Our central idea behind computing the variance of any $M_n$ in the aforementioned process $\{M_i\}_{i\geq 0}$ is to construct a martingale $(X_i, \CF_i)$ such that each $X_i$ is a linear function of $M_i$ (and $\CF_i$ is the canonical $\sigma$-algebra, $\sigma(X_0, \dots, X_i)$). For a martingale, computations of $\BE[X_n]$ and $\var[X_n]$ are more straightforward, and from them, we may obtain $\BE[M_n]$ and $\var[M_n]$. Our first step in developing this method is an elementary lemma that allows us to compute $\BE[M_n]$ for any $n$, when $\BE[M_{i+1} | M_i]$ is a non-constant linear function of $M_i$:

\lemma{Let $\{M_i\}_{i\geq 0}$ be a sequence of random variables with finite first moment, and let $M_0 = 0$ almost surely (a.s.). Let $\CF_i = \sigma(M_0, \dots, M_i)$ be the natural filtration. Suppose for all $i\geq 0$, we have $\BE[M_{i+1} | \CF_i] = a_iM_i + b_i$, where each $a_i$ and $b_i$ is fixed and known and each $a_i \neq 0$. Define
$$
q_i = \prod_{j=0}^i a_j,
$$
and let
$$
X_i = 
\frac{M_i}{q_{i-1}} - \sum_{j=0}^{i-1} \frac{b_j}{q_j}
$$
for $i\geq 1$. Set $X_0 = 0$ a.s.. Then $(X_i, \CF_i)$ is a martingale.
}

\begin{proof}
Clearly $X_i$ is $\CF_i$-adapted because $X_i$ is a deterministic function of $M_i$ and each $M_i$ is $\CF_i$-measurable. $\BE|X_i|$ is finite for all $i$ because $\BE|M_i|$ is finite for all $i$, and $X_i$ is a linear function of $M_i$. Finally, we compute that
\begin{align*}
    \BE[X_{i+1} | \CF_i] &= \frac{\BE[M_{i+1} | \CF_i]}{q_{i}} - \sum_{j=0}^{i} \frac{b_j}{q_j} \\
    &= \frac{a_iM_i + b_i }{q_{i}} - \sum_{j=0}^{i} \frac{b_j}{q_j} \\
    &= \frac{a_iM_i}{\prod_{j=0}^i a_j} + \frac{b_i}{q_i} - \sum_{j=0}^{i} \frac{b_j}{q_j} \\
    &= \frac{M_i}{\prod_{j=0}^{i-1} a_j}  - \sum_{j=0}^{i-1} \frac{b_j}{q_j} \\
    &= X_i
\end{align*} 
as desired.
\end{proof}

\corollary{For any $n > 0$, we have
$$
\BE[M_n] = q_{n-1}\sum_{j=0}^{n-1} \frac{b_j}{q_j}
$$.
}

\begin{proof}
    This follows immediately from the fact that
    $$
    0 = \BE[X_n] = \frac{\BE[M_n]}{q_{n-1}} - \sum_{j=0}^{n-1} \frac{b_j}{q_j}.
    $$
\end{proof}

\remark{
If $\BE[M_i^2]$ exists, then
$$
\var[M_i] = q_{i-1}^2 \var[X_i],
$$
a fact that will prove useful for computing $\var[M_n]$.
}

\lemma{Suppose we have a martingale $(X_i, \CF_i)$ such that $\BE[X_i^2]$ exists for all $i$, and $X_0 = 0$ a.s.. Suppose $\var[X_{i+1} | \CF_i] = \alpha_iX_i^2 + \beta_iX_i + \gamma_i$, where each $\alpha_i$, $\beta_i$, and $\gamma_i$ is fixed and known. Let $n > 0$. Then
$$
\var[X_n] = \sum_{i=0}^{n-1} \gamma_i \prod_{j=i+1}^{n-1} (\alpha_j + 1).
$$
Note that for $i=n-1$, we define the empty product as
$$
\prod_{j=n}^{n-1} (\alpha_j + 1) = 1.
$$
}

\begin{proof}
We proceed by induction. The base case $n=0$ holds trivially, as $\var[X_0] = 0$. For the inductive step, suppose the lemma holds when $n=k$ for some $k\geq 0$. Using the law of total variance and the fact that the unconditional expectation of any $X_i$ equals 0, we compute that
\begin{align*}
    \var[X_{k+1}] &= \BE[\var[X_{k+1} | \CF_k]] + \var[\BE[X_{k+1} | \CF_k]] \\
    &= \BE[\alpha_k X_k^2 +\beta_k X_k + \gamma_k] + \var[X_k] \\
    &= (\alpha_k + 1)\var[X_k] + \gamma_k \\
    &= (\alpha_k + 1)\left(\sum_{i=0}^{k-1} \gamma_i \prod_{j=i+1}^{k-1} (\alpha_j + 1)\right) + \gamma_k \\
    &= \left(\sum_{i=0}^{k-1} \gamma_i \prod_{j=i+1}^{k} (\alpha_j + 1)\right) + \gamma_k \\
    &= \sum_{i=0}^{k} \gamma_i \prod_{j=i+1}^{k} (\alpha_j + 1)
\end{align*}
as desired.

\end{proof}

\corollary{
Let $\{M_i\}_{i\geq 0}$ be a sequence of random variables with finite first and second moments, and let $M_0 = 0$ a.s.. Let $\CF_i = \sigma(M_0, \dots, M_i)$. Suppose $\BE[M_{i+1} | \CF_i] = a_iM_i + b_i$, where each $a_i$ and $b_i$ is fixed and known and each $a_i \neq 0$. Further, suppose $\var[M_{i+1} | \CF_i] = c_iM_i^2 + d_iM_i + e_i$, where each $c_i, d_i,$ and $e_i$ is fixed and known. Let
$$
s_i = \sum_{j=0}^i \frac{b_j}{q_j}.
$$
Then
$$
\var[M_n] = q_{n-1}^2\sum_{i=0}^{n-1} \left(\frac{c_iq_{i-1}^2s_{i-1}^2 + d_iq_{i-1}s_{i-1} + e_i}{q_i^2}\right) \prod_{j=i+1}^{n-1} \left( \frac{c_j}{a_j^2} + 1\right).
$$
}
\begin{proof}
Let $\{X_i\}_{i\geq 0}$ be as before. We have that
\begin{align*}
    \var[X_{i+1} | \CF_i] &= \frac{\var[M_{i+1} | \CF_i]}{q_i^2} \\
    &= \frac{c_iM_i^2 + d_iM_i + e_i}{q_i^2} \\
    &= \frac{1}{q_i^2}\left(c_iq_{i-1}^2 \left(X_i + s_{i-1}
    \right)^2 + d_iq_{i-1} \left(X_i + s_{i-1}
    \right) + e_i \right) \\
    &= \frac{c_iq_{i-1}^2}{q_i^2}X_i^2 + \frac{2c_iq_{i-1}^2s_{i-1} + d_iq_{i-1}}{q_i^2}X_i + \frac{c_iq_{i-1}^2s_{i-1}^2 + d_iq_{i-1}s_{i-1} + e_i}{q_i^2}
\end{align*}
Applying Lemma 2 with 
$$
\alpha_i = \frac{c_iq_{i-1}^2}{q_i^2} = \frac{c_i}{a_i^2} \quad \text{and} \quad \gamma_i = \frac{c_iq_{i-1}^2s_{i-1}^2 + d_iq_{i-1}s_{i-1} + e_i}{q_i^2}
$$
we obtain
$$
\var[X_n] = \sum_{i=0}^{n-1} \left(\frac{c_iq_{i-1}^2s_{i-1}^2 + d_iq_{i-1}s_{i-1} + e_i}{q_i^2}\right) \prod_{j=i+1}^{n-1} \left( \frac{c_j}{a_j^2} + 1\right),
$$
and thus
$$
\var[M_n] = q_{n-1}^2\sum_{i=0}^{n-1} \left(\frac{c_iq_{i-1}^2s_{i-1}^2 + d_iq_{i-1}s_{i-1} + e_i}{q_i^2}\right) \prod_{j=i+1}^{n-1} \left( \frac{c_j}{a_j^2} + 1\right)
$$
by Remark 1.
\end{proof}

\section{Relationship to Generalized Pólya Urns}

Lemmas 1 and 2 can help us understand the probability distribution associated with any generalized Pólya Urn scheme in which some quantity of interest $M_{i+1}$ at the $(i+1)$th time step has the property that $\BE[M_{i+1}|M_i] = a_iM_i + b_i$ for $a_i$ nonzero and $\var[M_{i+1}|M_i] = c_iM_i^2 + d_iM_i + e_i$. The number of white balls in the RPW model is one such quantity. More precisely, let $M_i$ denote the number of white balls that have been added to the urn after $i$ steps (with $M_0 = 0$ a.s.). As before, assume the urn starts with $u$ white balls and $v$ black balls. After $i$ steps, the urn contains $u + M_i$ white balls and $v + i - M_i$ black balls. Hence, conditional on $M_i$, the distribution of $M_{i+1}$ is given by
$$
M_{i+1}|M_i \sim M_i + \text{Bernoulli}\left((1-p_W) \left(\frac{u+M_i}{u+v+i}\right) + p_B\left(\frac{v+i-M_i}{u+v+i}\right)\right).
$$
Therefore,
$$
\BE[M_{i+1} | M_i] =  \left(1+\frac{1-p_B - p_W}{u+v+i}\right)M_i+\frac{p_B(v+i)+(1-p_W) u}{u+v+i}
$$
and, using the fact that a $\text{Bernoulli}(p)$ random variable has variance $p(1-p)$,
\begin{align*}
    \var[M_{i+1} | M_i] = &-\frac{\left(1 -p_B- p_W\right){}^2}{(u+v+i)^2}M_i^2 \\
    & + \frac{\left(p_B+p_W-1\right) \left(2 p_B (i+v)-i-2 u p_W+u-v\right)}{(u+v+i)^2} M_i \\
    & + \frac{\left(\left(1-p_B\right) (i+v)+u p_W\right) \left(p_B (i+v)+u \left(1-p_W\right)\right)}{(u+v+i)^2}.
\end{align*}
These calculations imply the following corollary:

\prop{
In the context of the RPW model, if we take $M_i$ to represent the number of white balls added to the urn after $i$ steps, the mean and variance of $M_n$ are given by Corollaries 1 and 2, respectively, when we set:
\begin{align*}
    a_i &= 1+\frac{1-p_B - p_W}{u+v+i} \\
    b_i &= \frac{p_B(v+i)+(1-p_W) u}{u+v+i} \\
    c_i &= -\frac{\left(1 -p_B- p_W\right){}^2}{(u+v+i)^2} \\
    d_i &= \frac{\left(p_B+p_W-1\right) \left(2 p_B (i+v)-i-2 u p_W+u-v\right)}{(u+v+i)^2} \\
    e_i &= \frac{\left(\left(1-p_B\right) (i+v)+u p_W\right) \left(p_B (i+v)+u \left(1-p_W\right)\right)}{(u+v+i)^2}.
\end{align*}
}
\begin{proof}
We need to verify that the conditions of Corollary 2 hold (which imply the conditions of Lemma 1 and therefore Corollary 1). By definition, $M_0 = 0$ a.s.. by definition. Since the support of $M_i$ is finite and equal to $\{0, 1, \dots, i\}$, all moments of $M_i$ are finite. The above calculations imply $\BE[M_{i+1} | \CF_i] = a_iM_i + b_i$ and $\var[M_{i+1} | \CF_i] = c_iM_i^2 + d_iM_i + e_i$, so all that remains to show is $a_i \neq 0$ for all $i$. Suppose for contradiction that $a_i = 0$ for some $i$. Rearranging, we obtain
$$
p_B + p_W - 1 = u + v + i.
$$
Since $u + v \geq 1$ (because the urn must start with at least one ball) and since $i$ is always a non-negative integer, the right-hand side is always at least 1. By assumption that $0 < p_B < 1$ and $0 < p_W < 1$, the left-hand side is always less than 1, giving us a contradiction. Therefore $a_i \neq 0$ for all $i$, allowing us to apply Corollaries 1 and 2 to compute the mean and variance of $M_n$.
\end{proof}

\section{Approximating the Probability Mass Function}

While Corollary 3 enables us to compute the exact variance of $M_n$ in the context of the RPW model, to the best of our knowledge, a non-recursive form for the entire probability mass function of $M_n$ has yet to be derived. Intuitively speaking, some of the difficulty in obtaining such a non-recursive form comes from the fact that color changes (i.e. steps in which we draw a black ball and add back a white ball or vice versa) in early steps impact $M_n$ much more drastically than in later steps, even though the asymptotic behavior of $M_n$ is deterministic: as Smythe and Rosenberger show,
$$
\frac{M_n}{n} \xrightarrow{\text{a.s.}} \frac{p_B}{p_W + p_B}
$$
as $n\to \infty$ (Smythe and Rosenberger 1995). That being said, Corollary 3 offers a means of approximating the entire probability mass function of $M_n$ when $p_W$ and $p_B$ are small and $n$ is large. For ease of interpretation, we focus on approximating the fraction $R_n = \frac{M_n}{n}$. The idea behind our approximation is that we choose some $0 < k < n$ such that the number of color changes occurring before step $k$ is small, with most of the probability mass at 0. At the same time, however, we want $k$ to be large enough such that no matter the composition of the urn after $k$ steps, the variance in the fraction of white balls $n-k$ steps later is small. Then, as we will show, the distribution of the number of white balls at step $k$, conditional on the step at which a white ball first appears, is approximately Beta-Binomial. A deterministic drift term then yields the approximate distribution of $R_n$, given the number of white balls at step $k$.

Let us formalize this intuition. Let $M_k$ be a random variable denoting the number of white balls added to the urn between steps 1 and $k$, and let $M_k^*$ denote an approximate version of $M_k$ (whose distribution we specify below). Assume without loss of generality that the urn starts out with either nonzero black balls and nonzero white balls ($u>0$ and $v>0$), or nonzero black balls and zero white balls ($u=0$ and $v>0$). (The case of the urn starting out with zero black balls and nonzero white balls is analogous to the case $u=0$, $v>0$). Our idea is to condition on the first step at which a white ball appears in the urn. Importantly, this step may be step 0, i.e. prior to adding any new balls to the urn, as in the case $u>0, v>0$. Using this idea, we construct a random variable $M_k^*$ to approximate $M_k$, whose distribution is as follows:
$$
\BP(M_k^* = x) = 
\begin{cases}
    f_{\text{Beta-Bin}}(x; k, u, v), & u>0, v>0 \\
    (1-p_B)^k, & u=0, v>0, x=0 \\
    \frac{\Gamma (k) (k+(v-1) x+v) \Gamma (k+v-x)}{x (x+1) \Gamma (k+v) \Gamma (k-x+1)}(1-(1-p_B)^k), & u=0, v>0, 0 < x \leq k,
\end{cases}
$$
where $f_{\text{Beta-Bin}}(x; k, u, v)$ denotes the Beta-Binomial probability mass function evaluated at $x$ with $k$ trials and Beta hyperparamters $u$ and $v$. The error of the PMF of $M_k^*$ as compared to the PMF of $M_k$ may be bounded as follows:
\prop{
When $u>0$ and $v>0$, we have
$$
\BP(M_k^* = x)(1-p_{\max})^k \leq \BP(M_k = x) \leq \BP(M_k^* = x)(1-p_{\min})^k + 1 - (1-p_{\max})^k,
$$
where $p_{\min} = \min(p_W, p_B)$ and $p_{\max} = \max(p_W, p_B)$. When $u=0$ and $v>0$, we have
$$
\BP(M_k^*=x)\left(\frac{kp_B(1-p_{\max})^{k-1}}{1-(1-p_B)^k}\right) \leq \BP(M_k = x)
$$
and
$$
\BP(M_k = x) \leq \BP(M_k^* = x)\left(\frac{kp_B(1-p_{\min})^{k-1}}{1-(1-p_B)^k}\right) + 1-(1-p_B)^k - kp_B(1-p_{\max})^{k-1},
$$
for $x > 0$. When $x = 0$, $\BP(M_k = x) = \BP(M_k^* = x)$.
}


\begin{proof}
The approximation in the case $u>0,v>0$ is given by
$$
M_k^* \sim \text{Beta-Bin}(k, u, v).
$$
To quantify the error of this approximation, we use the fact that in the absence of any color changes among the first $k$ steps, the distribution of $M_k$ is exactly $\text{Beta-Bin}(k, u, v)$. The probability of a transition occurring at any given step is bounded above by $p_{\max}$ and bounded below by $p_{\min}$. Hence the probability of zero color changes occurring in the first $k$ steps is bounded above by $(1-p)^k$. Letting $T$ be the number of color changes in the first $k$ steps, we have, for any $0\leq x \leq k$:
\begin{align*}
    \BP(M_k = x) &= \BP(M_k=x|T=0)\BP(T=0) + \BP(M_k=x|T>0)\BP(T>0) \\
    &\geq \BP(M_k^* = x)(1-p_{\max})^k
\end{align*}
and, for the bound in the other direction, we have
\begin{align*}
    \BP(M_k = x) \leq \BP(M_k^* = x)(1-p_{\min})^k + 1 - (1-p_{\max})^k.
\end{align*}
For the case $u=0, v>0$, once again let $T$ be the number of color changes in the first $k$ steps. Let $S$ denote the step at which a white ball first appears. Conditional on the event that exactly one transition occurs in the first $k$ steps, we have
$$
S|(T=1) \sim \text{Unif}\{1, \dots, k\}
$$
and
$$
M_k |(S, T=1) \sim 1+\text{Beta-Bin}(k-S, 1, v + S - 1)
$$
As one can compute, marginalizing $S$ yields
\begin{align*}
    \BP(M_k = x|T=1) &= \frac{1}{k}\sum_{s=1}^k f_\text{Beta-Bin}(x-1; k-S, 1, v + S - 1) \\
    &= \frac{\Gamma (k) (k+(v-1) x+v) \Gamma (k+v-x)}{x (x+1) \Gamma (k+v) \Gamma (k-x+1)}.
\end{align*}
Quantification of the error in this approximation proceeds similarly. For $x=0$, we have $\BP(M_k^* = x) = \BP(M_k=x)$, i.e. the error is 0. For $x>0$, we expand as follows:
\begin{align*}
    \BP(M_k = x) &= \BP(M_k=x|T=0)\BP(T=0) + \BP(M_k=x|T=1)\BP(T=1) + \BP(M_k=x|T>1)\BP(T>1).
\end{align*}
The first term on the right hand side is 0, as the event $M_k=x$ for $x>0$ is impossible in the event of 0 color changes. The second term we rewrite as follows:
\begin{align*}
    \BP(M_k=x|T=1)\BP(T=1) &= \BP(M_k=x|T=1)(\BP(T>0) - \BP(T>1)) \\
    &=  \BP(M_k=x|T=1)\BP(T>0)\left(1 - \frac{\BP(T>1)}{\BP(T>0)}\right) \\
    &=  \BP(M_k^*=x)\left(\frac{\BP(T=1)}{\BP(T>0)}\right) \\ 
    &\geq  \BP(M_k^*=x)\left(\frac{kp_B(1-p_{\max})^{k-1}}{1-(1-p_B)^k}\right).
\end{align*}
Therefore,
$$
\BP(M_k = x) \geq \BP(M_k^*=x)\left(\frac{kp_B(1-p_{\max})^{k-1}}{1-(1-p_B)^k}\right).
$$
We use a similar approach for the bound in the other direction. Expanding $\BP(M_k=x)$ in the same way, we have that
$$
\BP(M_k = x|T=1)\BP(T=1) \leq \BP(M_k^* = x)\left(\frac{kp_B(1-p_{\min})^{k-1}}{1-(1-p_B)^k}\right).
$$
Using the fact that
$$
\BP(T>1) \leq 1-(1-p_B)^k - kp_B(1-p_{\max})^{k-1},
$$
we conclude that
$$
\BP(M_k = x) \leq \BP(M_k^* = x)\left(\frac{kp_B(1-p_{\min})^{k-1}}{1-(1-p_B)^k}\right) + 1-(1-p_B)^k - kp_B(1-p_{\max})^{k-1}.
$$
This completes all cases and thus completes the proof.
\end{proof}

\remark{We expect tighter bounds if we assume at most $c$ color changes occur in the first $k$ steps, for some $c > 1$, and then condition on both the steps at which they occur as well as the type of color change, i.e. black-to-white or white-to black. We found the bounds given by Proposition 2 to be sufficient for the application to follow.}

\bigskip

\noindent We now move on to approximating $M_n$ and $R_n$, conditional on $M_k$. To do this, define a new Markov process $\{J_i\}_{i\geq 0}$ to represent the number of white balls added to the urn between steps $k$ and $k+i$. Set $u = M_k$ and $v = k-M_k$, and define
\begin{align*}
    a_i &= 1+\frac{1-p_B - p_W}{u+v+i} \\
    b_i &= \frac{p_B(v+i)+(1-p_W) u}{u+v+i} \\
    c_i &= -\frac{\left(1 -p_B- p_W\right){}^2}{(u+v+i)^2} \\
    d_i &= \frac{\left(p_B+p_W-1\right) \left(2 p_B (i+v)-i-2 u p_W+u-v\right)}{(u+v+i)^2} \\
    e_i &= \frac{\left(\left(1-p_B\right) (i+v)+u p_W\right) \left(p_B (i+v)+u \left(1-p_W\right)\right)}{(u+v+i)^2}
\end{align*}
as in Proposition 1 Set
$$
\mu_{n,k,M_k} = \BE[J_{n-k}|M_k] = q_{n-k-1}\sum_{j=0}^{n-k-1} \frac{b_j}{q_j}
$$
as in Corollary 1, where
$$
q_i = \prod_{j=0}^i a_j.
$$
Set
$$
\sigma_{n,k,M_k}^2 = \var[J_{n-k}|M_k] = q_{n-k-1}^2\sum_{i=0}^{n-k-1} \left(\frac{c_iq_{i-1}^2s_{i-1}^2 + d_iq_{i-1}s_{i-1} + e_i}{q_i^2}\right) \prod_{j=i+1}^{n-k-1} \left( \frac{c_j}{a_j^2} + 1\right)
$$
as in Corollary 2, where
$$
s_i = \sum_{j=0}^i \frac{b_j}{q_j}.
$$
We propose the approximation
$$
J_{n-k}|M_k = \mu_{n,k,M_k} \quad \text{a.s.}
$$
The error of this approximation, conditional on $M_k$, can be bounded probabilistically by Chebyshev's Inequality:
$$
\BP(|J_{n-k} - \mu_{n,k,M_k}| \geq t\sigma_{n,k,M_k}\,|\,M_k) \leq \frac{1}{t^2}
$$
for any real $t>0$. And, since $R_n = \frac{J_{n-k}+M_k}{n}$, we have $\var[R_n|M_k] = \frac{\sigma_{n,k,M_k}^2}{n^2}$; therefore
$$
\BP\left(\left|R_n - \frac{\mu_{n,k,M_k}}{n}\right| \geq \frac{t\sigma_{n,k,M_k}}{n}\,\Big|\,M_k\right) \leq \frac{1}{t^2}.
$$
To better understand the behavior of this Chebyshev bound, we propose the following upper bound for the quantity $\frac{\sigma_{n,k,M_k}^2}{n^2}$:
\prop{
Conditional on $M_k$, we have
$$
\frac{\sigma^2_{n,k,M_k}}{n^2} \leq \frac{1}{4k}.
$$
}
\begin{proof}
    Treat all random variables throughout this proof as being conditional on $M_k$. Set 
    $$
    X_i = \frac{J_i}{q_{i-1}} - \sum_{j=0}^{i-1} \frac{b_j}{q_j}
    $$
     as in Lemma 1. Conditional on $J_i$, $J_{i+1}$ is a binary random variable with support $\{J_i, J_i + 1\}$; hence, $\var[J_{i+1}|J_i]\leq \frac{1}{4}$. Transforming the $J_i$'s into the $X_i$'s tells us that
    $$
    \var[X_{i+1} | X_i] \leq \frac{1}{4q_{i}^2}.
    $$    
    Note that the right hand side does not depend on $X_i$. Hence we may apply Lemma 2 with $\alpha_i = \beta_i = 0$ and $\gamma_i = \frac{1}{4q_{i}^2}$ to obtain
    $$
    \var[X_{n-k}] \leq \sum_{i=0}^{n-k-1} \frac{1}{4q_{i}^2}.
    $$
    Therefore,
    \begin{align*}
        \var[J_{n-k}] &\leq q_{n-k-1}\sum_{i=0}^{n-k-1} \frac{1}{4q_{i}^2} = \frac{1}{4}\sum_{i=0}^{n-k-1} \prod_{j=i+1}^{n-k-1} a_j^2.
    \end{align*}
    The definition of $a_i$ in Proposition 1 tells us that
    $$
    a_j \leq 1 + \frac{1}{k+j}.
    $$
    Using the fact that
    $$
    \prod_{j=i+1}^{n-k-1} a_j^2 \leq \prod_{j=i+1}^{n-k-1}\left(1 + \frac{1}{k+j}\right)^2 = \left(\frac{n}{1+k+i}\right)^2,
    $$
    we obtain:
    \begin{align*}
        \var[J_{n-k}] &\leq \frac{n^2}{4}\sum_{i=0}^{n-k-1} \frac{1}{(1+k+i)^2} \\
        &\leq \frac{n^2}{4}\int_0^{n-k} \frac{dx}{(k+x)^2}\\
        &= \frac{n^2}{4}\left(\frac{1}{k}-\frac{1}{n}\right) \\
        &\leq \frac{n^2}{4k}.
    \end{align*}
    Thus, $\sigma^2_{n,k,M_k} \leq \frac{n^2}{4k}$. Dividing by $n^2$ proves the proposition.
\end{proof}

\section{Application to Genetics, with a Worked Example}

Propositions 2 and 3 quantify the error associated with the two steps we use to approximate the PMF of $R_n$, the fraction of white balls at step $n$. Here, we discuss and work through an application of this approximation to the genetics of viruses. Consider a single-stranded RNA virus, such as SARS-CoV-2. This strand is composed of a sequence of nucleotides (about 30,000 in the case of SARS-CoV-2), each of which may be adenine, cytosine, guanine, or uracil. When a host first contracts an infection, the number of copies of RNA in their body is small---typically about 1-2 particles for SARS-CoV-2 (Bendall et al. 2023). These copies of RNA then start to replicate, with each replication event sometimes introducing one or more substitutions, i.e. replacements of one nucleotide on the viral genome with another. If we focus our analysis on one particular site, or position along the viral genome, then we can take black balls in the RPW model to represent copies of RNA in which the site exhibits a certain nucleotide---say, adenine, for example---and white balls to represent copies of RNA in which the site exhibits any of the three other possible nucleotides (we will not distinguish between the three in the context of this application). Then we may take $p_B$ to represent the probability of an adenine nucleotide mutating into a cytosine, guanine, or uracil nucleotide in a replication event, and $p_W$ to represent the probability of a cytosine, guanine, or uracil nucleotide mutating into an adenine nucleotide. Under the Jukes-Cantor model of evolution, which assumes that each possible substitution of one nucleotide for a different nucleotide is equally likely, it follows that $p_B = 3p_W$ (Erickson 2010). Assuming the host is initially infected with a total of $u+v$ viral particles, we may take $v$ to represent the number with adenine at our site of interest and $u$ the number with any other nucleotide at our site of interest. (Our choice of adenine here was arbitrary; more generally, we may take $p_B$ to be the mutation rate from any chosen nucleotide to some other nucleotide, and $p_W$ the opposite. In the case of $u=0$ and $v=1$, which we discuss later on, it is convenient to make this chosen nucleotide whatever nucleotide the one original viral particle features at our site of interest.)

To see the relationship between the RPW model and viral replication more explicitly, consider the following pure-birth process: Suppose we start with $u$ white viral particles and $v$ black viral particles at time 0, where, as above, black viral particles exhibit some given nucleotide at some given site of interest, and white viral particles exhibit any other nucleotide at the same site of interest. Suppose each viral particle (both the original $u+v$ particles as well as any particle born later) replicates to form new viral particles over time, such that the times at which the offspring of a given particle are born follow a Poisson process with rate $\lambda$. Finally, assume that for each birth event, the probabilities of a white or black offspring equal $p_B$ and $1-p_B$ when the parent is black, respectively; and equal $1-p_W$ and $p_W$ when the parent is white, respectively. The previous three statements fully define a stochastic pure-birth model with mutations, but for a more intuitive sense of the overall growth process, we may equivalently say that if $N(t)$ is the number of particles at time $t$, then for small $\Delta_t$, we have
$$
N(t + \Delta_t)\,|\, N(t) \sim N(t) + \text{Bernoulli}(N(t)\lambda \Delta_t)).
$$
This follows from the fact that the number of birth events from $N(t)$ particles between time $t$ and time $t + \Delta_t$ is the sum of $N(t)$ independent $\text{Poisson}(\lambda \Delta_t)$ random variables, which follows a $\text{Poisson}(N(t)\lambda \Delta_t)$ distribution. As $\Delta_t \to 0$, this distribution approximates the $\text{Bernoulli}(N(t)\lambda \Delta_t))$ distribution as per the above.

We claim that the distribution of $W_n$, the number of white particles in this stochastic pure-birth model immediately following the $n$th birth, is equal in distribution to $M_n$ in the context of the RPW model. Since the color of the $n$th particle to be born conditional on its ancestor clearly follows the same conditional distribution as in the RPW model, it suffices to show that the ancestor of the $n$th particle to be born in the pure-birth model is uniformly distributed over the $u+v+n-1$ already-born particles. To see this, suppose that the ($n-1$)th particle in the pure-birth model is born at some time $t$. Since the births of any given particle follow a Poisson process, the times at which each of the $u+v+n-1$ particles present at time $t$ will next give birth are i.i.d $\text{Expo}(\lambda)$ random variables by the memoryless property. Hence, the first of these $u+v+n-1$ to replicate after time $t$ is uniformly distributed over the $u+v+n-1$ particles.


Now that we have demonstrated the applicability of the RPW model to pure-birth processes, we next show that our approximation in Section 4 fits the true RPW process reasonably well for values of $p_W, p_B$, and $n$ that reflect infectious diseases. We then apply our approximate PMF to real genetic sequencing data from viruses. For this example, we look at SARS-CoV-2, thanks to the wealth of data and scientific studies on the virus from the past several years. Bar-On et al.'s 2020 paper estimates $p$, the probability of one nucleotide mutating into any other nucleotide, to be $10^{-6}$ (Bar-On et al. 2020). Hence, we set $p_B = 10^{-6}$ and $p_W = p_B/3$, assuming the Jukes-Cantor model. In the context of our application, $n$ represents the total number of ``infectious units'' or viral particles capable of causing infection, which Bar-On et al. estimate to be on the order of $10^5$ to $10^8$; here, we set $n= 10^6$ (Bar-On et al. 2020). We further assume that the host was infected with a single viral particle (Bendall et al. 2023); therefore, we set $u = 0$ and $v = 1$. For $k$, we propose the heuristic $k = p^{-2/3} = 10^4$, which yields $kp = 10^{-2}$ expected color changes in the first $k$ steps---making the probability of two or more color changes before the $k$th step highly improbable. Per Propositions 2 and 3, we obtain the following error bounds from these values:
$$
0.995\times\BP(M_k^* = x) \leq \BP(M_k = x) \leq 1.002\times\BP(M_k^* = x) + 0.0000497
$$
$$
\BP\left(\left|R_n - \frac{\mu_{n,k,M_k}}{n}\right| \geq \frac{t}{200}\,\Big|\,M_k\right) \leq \frac{1}{t^2}, \quad \forall t>0
$$
where, as the reader may recall, $M_k$ is the number of mutated particles at step $k$, and $M_k^*$ is the approximate version of $M_k$ whose distribution is specified in Section 4. Another, perhaps more intuitive means of understanding the behavior of the process from step $k$ onwards is that conditional on $M_k$, $R_n = M_n/n$ is a random variable with mean $\frac{\mu_{n,k,M_k}}{n}$ and variance less than $\frac{1}{4k} = \frac{1}{4\times 10^4}$. Hence, $R_n$ takes on values approximately equal to $\frac{\mu_{n,k,M_k}}{n}$ with high probability.

To visualize this approximation, we simulated $10^6$ replications of the true Pólya Urn process, then plotted a histogram of $R_n$ against the approximation derived in Section 4 (see Figure 1). For the sake of visualization, we truncated the data to values of $R_n$ ranging from 0 to 0.002, and plotted the log density for both the histogram and the approximate PMF. Note that higher values on the x-axis have fewer observations; hence, the quality of the fit decreases as the proportion of white balls increases.

\begin{figure}[h]
\centering
\includegraphics[width=\textwidth]{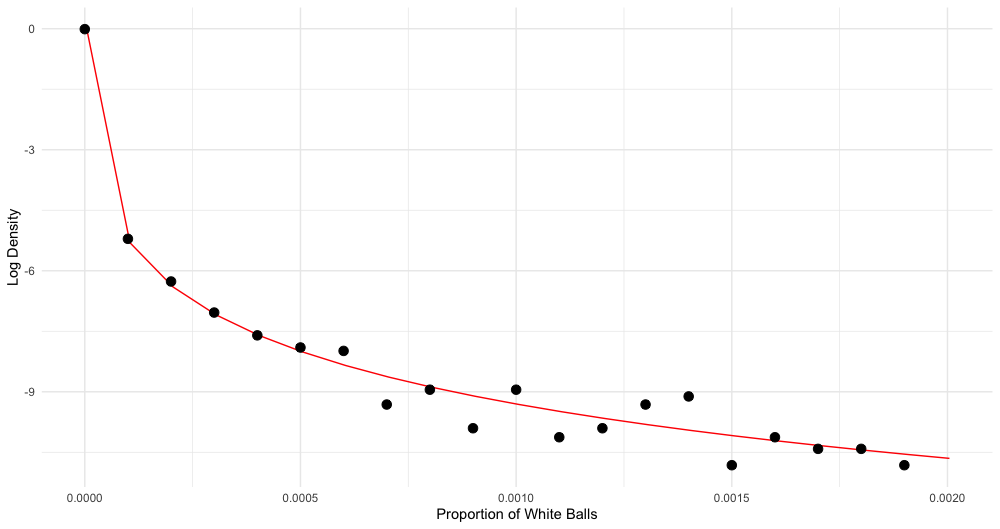}
\caption{Simulated log density (dots) and approximate log density (line) for the RPW model with $p_B=10^{-6}, p_W = \frac{1}{3} \times 10^{-6}, n=10^6, u=0, v=1, k=10^4$.}
\end{figure}

Finally, we measured the ability of our approximate density function for the RPW process to fit actual genetic data (see Figure 2). Our dataset consists of 40 PCR swabs collected from a 2020 SARS-CoV-2 outbreak in Grand Junction, Colorado, and sequenced at the Broad Institute of MIT and Harvard. For each sample, we treated each of the 29,903 sites on the SARS-CoV-2 genome as an independent realization of the RPW process. We omitted observations with a strand bias exceeding 10 or a total read depth of less than 1,000 (Guo et al. 2012). We fit our approximate density function using least-squares. The MLE of $p_B$ was $5.24 \times 10^{-6}$---equal in order of magnitude to the value proposed in Bar-On et al. (2020).

\begin{figure}[h]
\centering
\includegraphics[width=\textwidth]{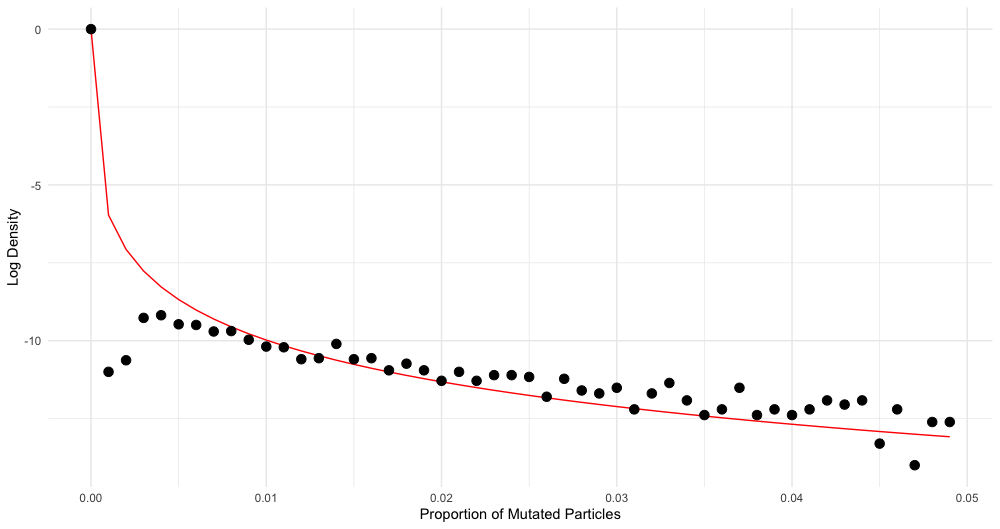}
\caption{Log density of the fraction of mutated particles in 40 individuals across 29,903 sites per individual (dots) and fitted log density (line). Note that the two apparent outliers near the y-axis come from the fact that the proportion of mutated particles is measured as the number of reads exhibiting a given SNP divided by the total number of reads. Since the total number of reads may vary, with a typical range being about 100 to 30,000, within-host variants with a very low true frequency will be underrepresented in sequencing data as they are more likely to go undetected.}
\end{figure}

\section{Conclusion}

The RPW model is a useful tool for capturing numerous phenomena in genetics, randomized control trials, and more. In this paper, we derive a method to rewrite the Pólya Urn model as a martingale, which facilitates computations of its mean and variance after any finite number of steps. Notably, this approach corrects an earlier attempt to compute the variance of the RPW process, which we show to be erroneous. Finally, we use our mean and variance formulae to derive an approximation of the density function of the RPW process when the transition probabilities $p_W$ and $p_B$ are small. By fitting this approximation to both simulated Pólya Urn data and real genetic data, we demonstrate the quality of the approximation as well as the applicability of the RPW model to the life sciences.

Understanding the probability distribution of viral particles within a host exhibiting a given mutation provides several new insights in viral genetics. First and foremost, it offers a novel method of estimating the viral mutation rate. Typically, any reliable estimation of this quantity requires observing the virus over long periods of time across numerous patients, then constructing phylogenetic trees and analyzing the mutations present in each branch. This method neglects the important relationship between within-host variant frequencies and the mutation rate---which, as shown above, allows for such an estimation using far fewer patient samples. In addition, the distribution of within-host variant frequencies can help us better understand viral transmission and evolution, as the passage of a low-frequency within-host variant from one host to another may lead to that variant becoming overrepresented in the population. Moving forward, we hope to incorprate this model into methods for viral transmission analysis, as well as demonstrate its ability to estimate the mutation rates of other viruses.

From a more theoretical standpoint, the RPW model is just one example of a generalized Pólya Urn model, and for future studies, we would like to explore what other generalized Pólya Urn models may be studied using martingales. A simple extension of the RPW model allows for the possibility of adding some $n>1$ balls (all of the same color) at each step (see Matthews and Rosenberger 1997). This model also satisfies the conditions of Lemmas 1 and 2 and thus may be rephrased as a martingale in similar fashion. Generalizations to multiple colors, however, may prove more challenging. For instance: if we let $\{M_i\}_{i\geq 0}$ denote the number of white balls in the urn as before, but now the urn may contain white, black, and red balls, then the expectation $\BE[M_{i+1} | \CF_i]$ (where $\CF_i$ denotes the $\sigma$-algebra generated by the first $i$ steps) now depends not only on $M_i$, but also on the breakdown of black and red balls at step $i$---assuming the black-to-white and red-to-white transition probabilities are not equal.

Overall, this paper underscores the power of analyzing stochastic processes using martingales. It moreover provides a key example of how such analyses may be applied to important questions in viral genetics. Moving forward, we hope to explore further both the theory and applications of models that may converted to martingales via similar processes.

\section*{Acknowledgements}
We would like to thank Patrick Varilly and Pardis C. Sabeti for their additional contributions to this research, as well as the Sabeti Lab at the Broad Institute of MIT and Harvard for providing the genomic sequencing data to which we applied our model. 

\section*{Funding}
Michael Mitzenmacher was supported in part by NSF grants CCF-2101140, CNS-2107078, and DMS-2023528.

\section*{Conflicts of Interest}
The authors report there are no competing interests to declare.

\section*{Appendix}

To the best of our knowledge, only one previous paper---Matthews and Rosenberger (1997)---has attempted to compute the variance of the RPW model. Matthews and Rosenberger consider a sligntly more general formulation of the urn problem: rather than adding one new ball to the urn at each step, they add some positive integer $\beta$ particles at each step, all of the same color. The version of the RPW model described in this paper is identical to that in Mathews and Rosenberger (1997) when we set $\beta = 1$. Their main result (with variables changed to reflect our notation, and with $\beta = 1$) states that:
\begin{align*}
    \var[M_n] =&\; 2t_Wt_B \sum_{l=1}^{n-2} \frac{\Gamma(c+l)}{\Gamma(c+2\lambda + l)} \sum_{k=l+1}^{n-1}\frac{\Gamma(c+2\lambda + k -1)}{\Gamma(c+k+\lambda)}\sum_{j=k}^{n-1} \frac{\Gamma(c+j+\lambda)}{\Gamma(c+j+1)} \\
    &+ 2\lambda(t_W - t_B)(vt_W - ut_B) \frac{\Gamma(c)}{\Gamma(c+\lambda)} \\
    &\quad \times \sum_{l=1}^{n-2} \frac{\Gamma(c+\lambda + l - 1)}{\Gamma(c + 2\lambda + l)}\sum_{k=l+1}^{n-1} \frac{\Gamma(c + 2\lambda + k - 1)}{\Gamma(c+k+\lambda)}\sum_{j=k}^{n-1} \frac{\Gamma(c + j + \lambda)}{\Gamma(c+j+1)} \\
    &- \left((vt_W - ut_B)\frac{\Gamma(c)}{\Gamma(c+\lambda)}\sum_{j=0}^{n-1} \frac{\Gamma(c+j+\lambda)}{\Gamma(c+j+1)}\right)^2 \\
    &+ nt_Wt_B + 2\lambda t_Wt_B \sum_{k=1}^{n-1} \frac{\Gamma(c+k)}{\Gamma(c+k+\lambda)}\sum_{j=k}^{n-1} \frac{\Gamma(c+j+\lambda)}{\Gamma(c+j+1)} \\
    &+ (t_W - t_B)(vt_W - ut_B) \frac{\Gamma(c)}{\Gamma(c+\lambda)}\sum_{k=1}^{n}\frac{\Gamma(c+\lambda + k-1)}{\Gamma(c+k)} \\
    &+ 2\lambda (t_W - t_B)(vt_W - ut_B) \frac{\Gamma(c)}{\Gamma(c+\lambda)}\sum_{k=1}^{n-1} \frac{\Gamma(c+\lambda+k-1)}{\Gamma(c+k+\lambda)}\sum_{j=k}^{n-1} \frac{\Gamma(c+\lambda+j)}{\Gamma(c+j+1)} \\
    &+ 2(vt_W - ut_B)^2 \frac{\Gamma(c)}{\Gamma(c+2\lambda)}\sum_{j=1}^{n-1} \frac{\Gamma(c+\lambda + j)}{\Gamma(c+j+1)}\sum_{k=1}^j \frac{\Gamma(c+2\lambda + k - 1)}{\Gamma(c + \lambda + k)}
\end{align*}
where
\begin{align*}
    t_W &= \frac{p_B}{p_B + p_W} \\
    t_B &= \frac{p_W}{p_B + p_W} \\
    c &= u+v \\
    \lambda &= 1 - p_W - p_B.
\end{align*}
Note that this variance expression includes corrections made by Matthews and Rosenberger on page 240, which is why it does not perfectly match the expression on page 237.

In a later paper, Rosenbuger evaluates this variance formula for certain combinations of input parameters (Rosenberger 1999). In particular, he claims that for $p_W = p_B = 0.5$, $u=v=1$, and $n=25$, we have
$$
\text{SD}\left[\frac{M_n}{n}\right] = \frac{\sqrt{\var[M_n]}}{n} \approx 0.139.
$$
In this special case, however, each ball added to the urn is an i.i.d. $\text{Binomial}(\frac{1}{2})$ draw. Hence, $M_n \sim \text{Binomial}(n, \frac{1}{2})$, so we should obtain
$$
\text{SD}\left[\frac{M_n}{n}\right] = \frac{1}{2\sqrt{n}} = 0.1.
$$
To help determine the source of error in the Matthews and Rosenberger paper, we observe that in the case $u=v=1$ and $p_B = p_W = \frac{1}{2}$, we have $t_W = t_B = \frac{1}{2}$, $\lambda = 0$, and $c=2$. Hence, the Matthews and Rosenberger variance expression simplifies to:
\begin{align*}
    \var[M_n] &= \left(2t_Wt_B \sum_{l=1}^{n-2} \frac{\Gamma(c+l)}{\Gamma(c+ l)} \sum_{k=l+1}^{n-1}\frac{\Gamma(c+ k -1)}{\Gamma(c+k)}\sum_{j=k}^{n-1} \frac{\Gamma(c+j)}{\Gamma(c+j+1)}\right)+ nt_Wt_B  \\
    &= \left(\frac{1}{2}\sum_{l=1}^{n-2} \sum_{k=l+1}^{n-1}\frac{1}{c+k-1}\sum_{j=k}^{n-1} \frac{1}{c+j}\right) + \frac{n}{4}
\end{align*}
Since the correct variance in this case equals $\frac{n}{4}$, the first summand in Matthews and Rosenberger's variance expression appears to be the source of error.

%


\newpage


\begin{thebibliography}{9}

\bibitem{wd}
Wei, L.J. \& Durham, S. The Randomized Play-the-Winner Rule in Medical Trials. \textit{Journal of the American Statistical Association} \textbf{73}, 364 (1978). 

\bibitem{sr}
Smythe, R.T. \& Rosenberger, W.F. Play-the-Winner Designs, Generalized Pólya Urns, and Markov Branching Processes. \textit{IMS Lecture Notes - Monograph Series} \textbf{25} (1995).

\bibitem{s}
Smythe, R.T. Central limit theorems for urn models. \textit{Stochastic Processes and their Applications} \textbf{65} (1996).

\bibitem{rs}
Rosenberger, W.F. \& Sriram, T.N. Estimation for an adaptive allocation design. Journal of Statistical Planning and Inference \textbf{59}, 2 (1997).

\bibitem{b}
Bendall, E.E., Callear, A.P., Getz, A. et al. Rapid transmission and tight bottlenecks constrain the evolution of highly transmissible SARS-CoV-2 variants. \textit{Nature Communications} 14, 272 (2023).

\bibitem{jc}
Erickson, K. (2010). The Jukes-Cantor Model of Molecular Evolution. \textit{Primus}, \textbf{20}, 5 (2010).

\bibitem{se}
Bar-On, Y.M., Flamholz, A., Phillips, R., \& Milo, R. SARS-CoV-2 (COVID-19) by the numbers. \textit{eLife} \textbf{9}, e57309 (2020).

\bibitem{g}
Guo, Y., Li, J., Li, CI. et al. The effect of strand bias in Illumina short-read sequencing data. BMC Genomics 13, 666 (2012).

\bibitem{kn}
Knape, M. \& Neininger, R. Pólya Urns Via the Contraction Method. \textit{Combinatorics, Probability and Computing} \textbf{23}, 6 (2014).

\bibitem{mr}
Matthews, P.C. \& Rosenberger, W.F. Variance in randomized play-the-winner clinical trials. \textit{Statistics \& Probability Letters} \textbf{35} (1997).

\bibitem{r}
Rosenberger, W.F. Randomized play-the-winner clinical trials: review and recommendations. \textit{Control Clinical Trials} \textbf{20}, 4 (1999).






\end{thebibliography}
\end{document}